\documentclass[10pt,letterpaper,twocolumn]{article} 

\usepackage{ol2}
\usepackage[draft]{hyperref}
\usepackage{amsmath}

\begin{document}

\twocolumn[ 

\title{Anderson localization in a periodic photonic lattice\\ with a disordered boundary}

\author{U. Naether,$^{1,2,3,*}$ J. M. Meyer,$^3$ S. St\"{u}tzer,$^3$  A. T\"{u}nnermann,$^3$ S. Nolte,$^3$ M. I. Molina,$^{1,2}$  and A. Szameit$^{3}$}

\address{
$^1$Departamento de F\'isica, Facultad de Ciencias, Universidad de
Chile, Santiago, Chile
\\
$^2$Center for Optics and Photonics (CEFOP), Casilla 4016,
Concepti\'on, Chile \\
$^3$University Jena, Institute of Applied Physics, D-07743 Jena, Germany \\
$^*$Corresponding author: unaether@u.uchile.cl}

\begin{abstract}
We investigate experimentally the light evolution inside a
two-dimensional finite periodic array of weakly-coupled optical
waveguides with a disordered boundary. For a completely localized
initial condition away from the surface, we find that the disordered
boundary induces an asymptotic localization in the bulk, centered
around the initial position of the input beam.
\end{abstract}

\ocis{130.2790, 240.6700, 290.5825}

 ] 

\noindent The original concept of Anderson localization assumes a
periodic structure where disorder is introduced via a random
change of the local properties at each site of the lattice, leading to wave localization due to interference between multiple scattering paths
\cite{al}. Although it was first described in the context of
condensed matter\cite{al,abr}, there are now examples in many other
fields: Acoustics\cite{acoustics}, microwaves\cite{mw},
Bose-Einstein condensates\cite{bec} and optics \cite{john,
freiopt,wiersma,exp2d,exp1d}, to name a few. Most importantly, in the
optical domain it was proven in experiment and theory that a random
displacement of the lattice sites yields the same localization
results as changing the local properties of the individual sites
\cite{sand,and}.

In rough or corrugated channels, the phenomenon of Anderson localization is connected to the
transmission of electrons or optical pulses. In these systems, the channel
surface is disordered along the propagation direction. Theoretical
studies showed transitions between diffusive and localized regimes
\cite{garcia1,garcia2,garcia3}. Transport behavior was also
considered in\cite{freil1, freil2,freil3}, where multiple scattering
from longitudinal surface roughness caused localization of waves,
although regimes of coexistence between ballistic, diffusive and
localized transport, depending on the symmetry of the corrugation
profile, have been predicted\cite{izrailev}.

Recently, a different kind of ``corrugated waveguide'' was
considered theoretically \cite{mariopla}. Here the disorder is only in the
transverse direction, and does not change along the direction of
propagation. Also, the optical medium in the bulk, away from the
corrugated surface, possesses a periodic index of refraction along
the two transversal directions, thus forming an array of weakly
coupled waveguides \cite{dlb}. The disorder is imposed on the
boundary by random displacement. Despite the weak disorder, a light
beam still tends to localize in the center of such a system far from
the boundary. In our work, we experimentally prove this prediction.
\begin{figure}[t]
\centering
\includegraphics[width=0.4\textwidth]{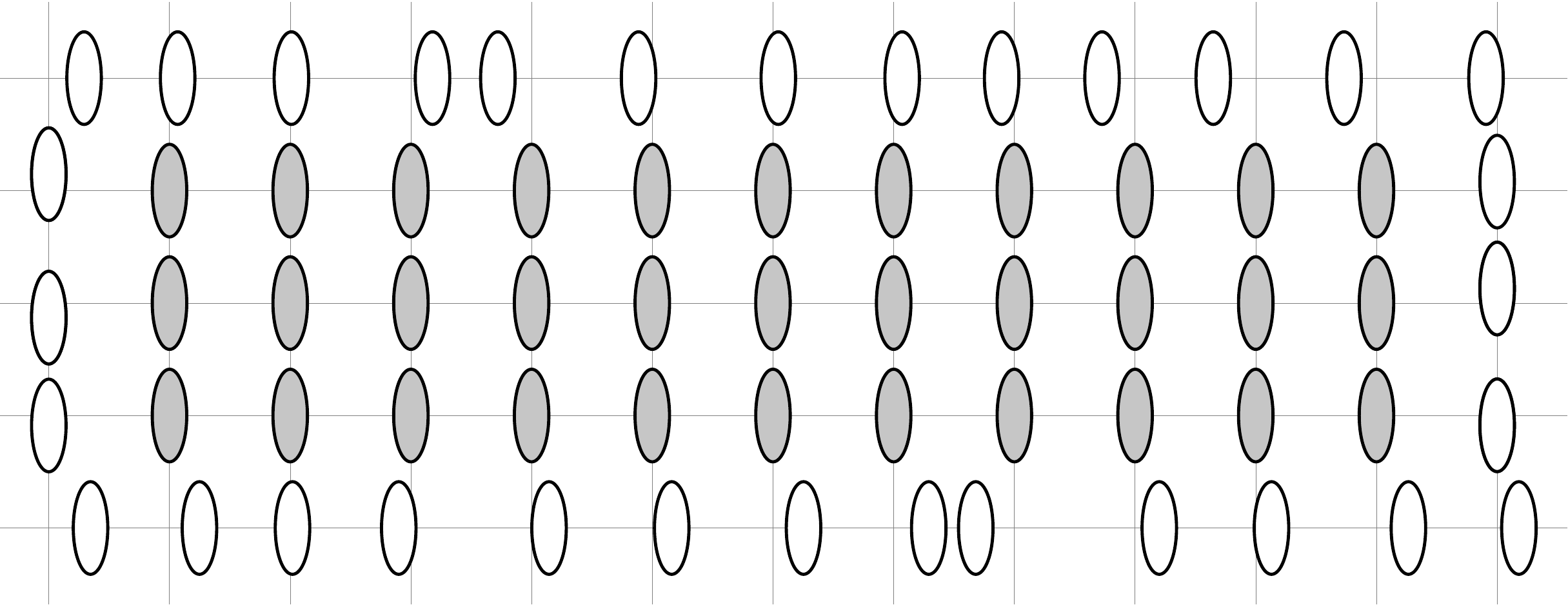}
\caption{Scheme of a finite two-dimensional coupled array of
elliptical waveguides with disordered boundary.} \label{scheme}
\end{figure}

Let us consider a 2D rectangular $N\times M$ waveguide array
(Fig.\ref{scheme}). In the coupled-mode approach, the electric field
$E({\bf r},z)$ propagating along the waveguides can be written as a
superposition of the waveguide modes, $u({\bf r},z)=\sum_{\bf n}
u_{\bf n}(z) \phi({\bf r}-{\bf n})$, where ${\bf r}=(x,y)$, $u_{\bf
n}$ is the amplitude of the single modes $\phi_{\bf n}$ centered
around site ${\bf n}=(p,q)$. The evolution equations for the mode
amplitudes $u_{\bf n}$ are
\begin{eqnarray}
\sum_{j=p\pm1}C_{v_j} u_{j,q} +\sum_{j=q\pm1}C_{h_j} u_{p,j} =
-i{du_{pq}\over{dz}} \; , \label{ode}
\end{eqnarray}
where ${\bf n}=(p,q)$ denotes the position of the guide center, $z$
is the longitudinal distance, and the $C_{v,h}$ are the coupling coefficients
between nearest-neighbors guides. They decay exponentially with the
mutual distance between the guides. To keep our approach general, we
assume anisotropic coupling in the horizontal and vertical
direction. At the boundary of the array, randomness is introduced by
displacing the guides from their (ordered) positions, along the
boundary surface. This creates random couplings among the boundary
guides, for the horizontal coupling $ C_{h}\rightarrow C_{h}
e^{-\omega_{h} \Delta}$ as well as for the vertical coupling $
C_{v}\rightarrow C_{v} e^{-\omega_{v} \Delta}$. The quantity
$\Delta$ is a random number in $[-1,1]$ and the randomness strengths
$\omega_{h}$ and $\omega_{v}$ along the horizontal and vertical
boundaries respectively are different due to the ellipticity of the
guides. The bulk, that represents the ordered part of the array, is
connected to the disordered boundary layer with coupling values
computed from the mean pythagorean distance. These conditions are
close to the possibilities of experimental realization\cite{and}.
The entire system could be viewed as a single ``corrugated''
photonic crystal waveguide, with fixed transversal corrugation.

\begin{figure}[t]
\centering
\includegraphics[width=0.45\textwidth]{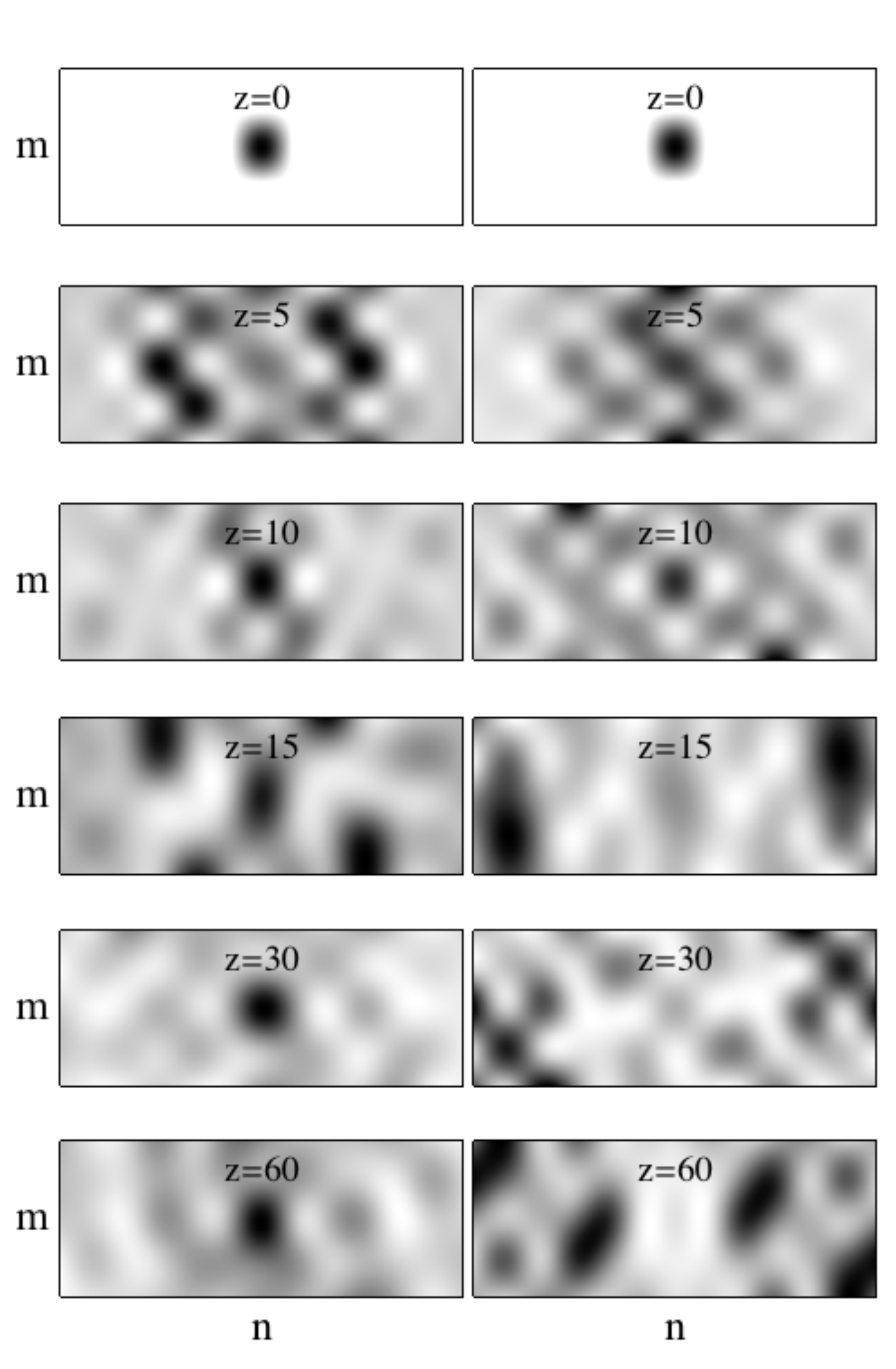}
\caption{Comparison between simulations of an ordered $13\times5$
array (right column), and the disordered boundary waveguide
array)of eq. (\ref{ode}) (left column) with $C_{\alpha}=2$,
$C_{\beta}=1.7$, $\omega_{\alpha}=11/28$, $\omega_{\beta}=4/17$,
averaged over 100 realizations.} \label{sim13}
\end{figure}

First, in order to illustrate the localization mechanisms in the
system under consideration, we analyze a waveguide array with $13
\times 5$ waveguides and disordered boundary and perform a direct
numerical integration of Eq. (\ref{ode}) with a single-site
excitation at the array center $(n_c,m_c)=(7,3)$. The parameters
chosen for the simulation of the disordered case (left column of
Fig. \ref{sim13}) are: $C_{h}=2$, $C_{v}=1.7$, $\omega_{h}=11/28$,
$\omega_{v}=4/17$ \cite{coupl}. For comparison, we compute the
evolution for the ordered case and its pure discrete diffraction
(right column of Fig. \ref{sim13}). For small propagation distances
($z\simeq5$), both systems show discrete diffraction with maxima at
the outer lobes. The amplitude of the initially excited site
decreases steadily with $z$. However, after a transition distance of
$z_t\simeq10$, that is necessary for a complete backscattering
cycle, the influence of the boundary layer becomes apparent. In the
case of the disordered boundary, for $z>10$, the amplitude of the center
site is non-vanishing for all propagation lengths. It continues to
oscillate with $z$, but finally saturates to a nearly stationary
mode, showing the persistence and stability of localization due to
the disordered boundary. The main difference to the case of a
completely disordered system (bulk + surface) is that the initial
oscillations due to discrete diffraction are stronger.

For our experiments, we fabricated various waveguide arrays in
polished bulk fused silica glass, using the laser direct-writing
technology\cite{arrays}. Each guide has dimensions of $4\times12$
$\mu \mbox{m}^2$ and exhibits a refractive index change of $\approx
5\times 10^{-4}$\cite{coupl}. We prepared $60$ disordered-boundary
waveguide arrays with a length of $101$  $\mbox{mm}$, with either
$N\times M=5\times5$ or with $N\times M=13\times5$ waveguides each.
The inter-guide separation in the ordered bulk, which is also the
mean separation at the (disordered) boundary, was of $14$ $\mu
\mbox{m}$ and $17$ $\mu \mbox{m}$, along the horizontal and vertical
directions, respectively. Disorder in the waveguide spacing was
induced by varying the inter-guide distance: $d_{h}=14\pm 5.5\Delta$
$\mu \mbox{m}$ and $d_{v}=17\pm 4\Delta$ $\mu \mbox{m}$, with $\Delta\in [-1,1]$ randomly equally distributed.

In each array the individual central waveguide was excited using a
Ti:Sapphire laser system at low input power to ensure linear
propagation. At the end facet, the intensity patterns were recorded
with a CCD camera. For the observation of localization in the
$5\times5 $ arrays, a wavelength of $800$ $\mbox{nm}$ was used,
corresponding to coupling coefficients\cite{coupl} of
$C_{h}(C_{v})\simeq 2.0 (1.7)$ $\mbox{cm}^{-1}$. To enhance
localization in the arrays of $13\times5$ waveguides, the wavelength
was increased to $840\ \mbox{nm}$, leading to higher coupling
coefficients and therefore decreased coupling lengths. This allows
the observation of localization effects at effectively shorter
evolution distances.

\begin{figure}[t]
\centering
\includegraphics[width=0.46\textwidth]{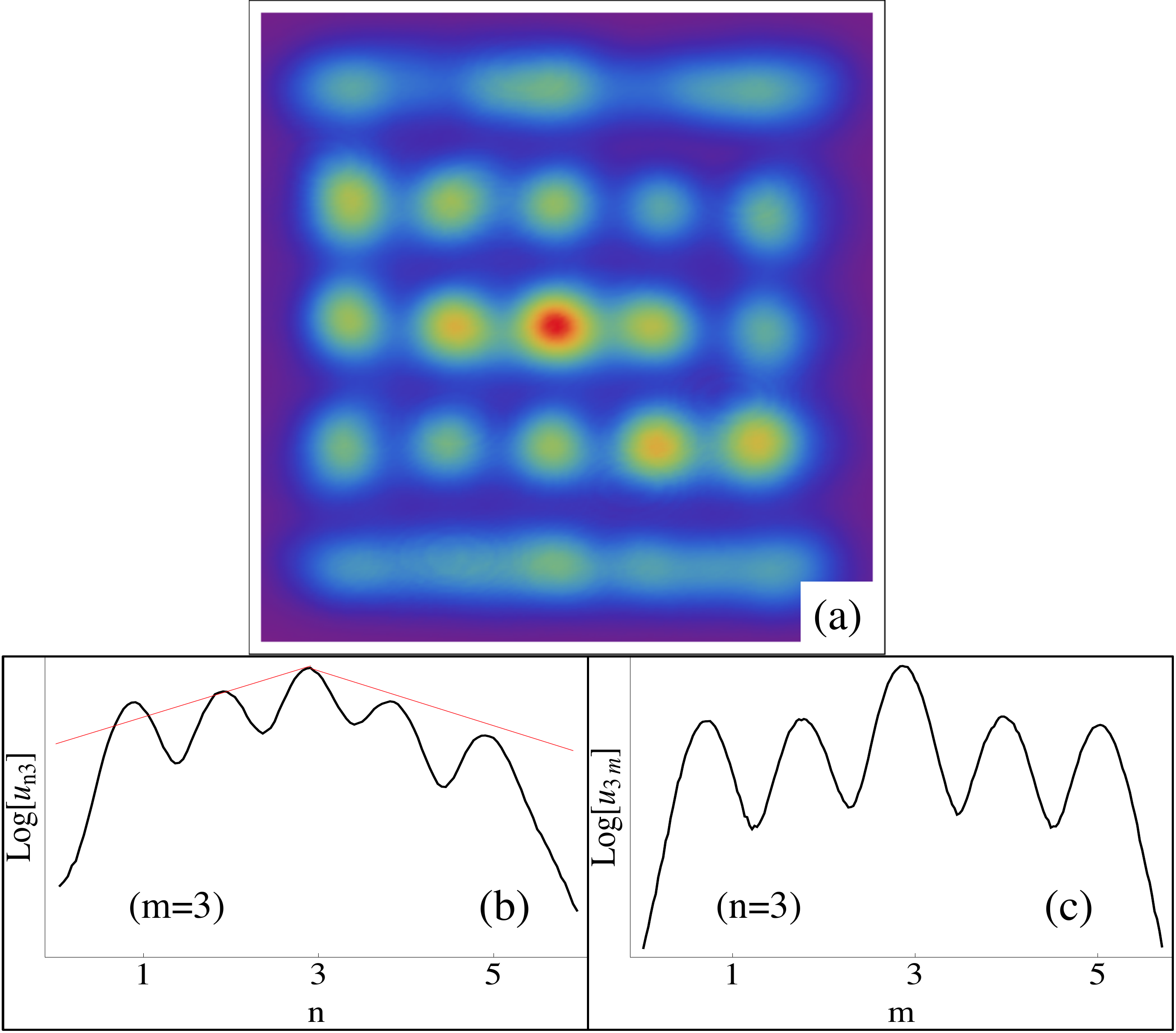}
\caption{Experimental results for the $5\times5$ arrays: (a) mean
output of 30 realizations, $\lambda=800$ $nm$, (b), (c) logarithmic
plots of $|u_{n,m_c}|$, $|u_{n_c,m}|$ respectively.} \label{5x5}
\end{figure}

\begin{figure}[t]
\centering
\includegraphics[width=0.46\textwidth]{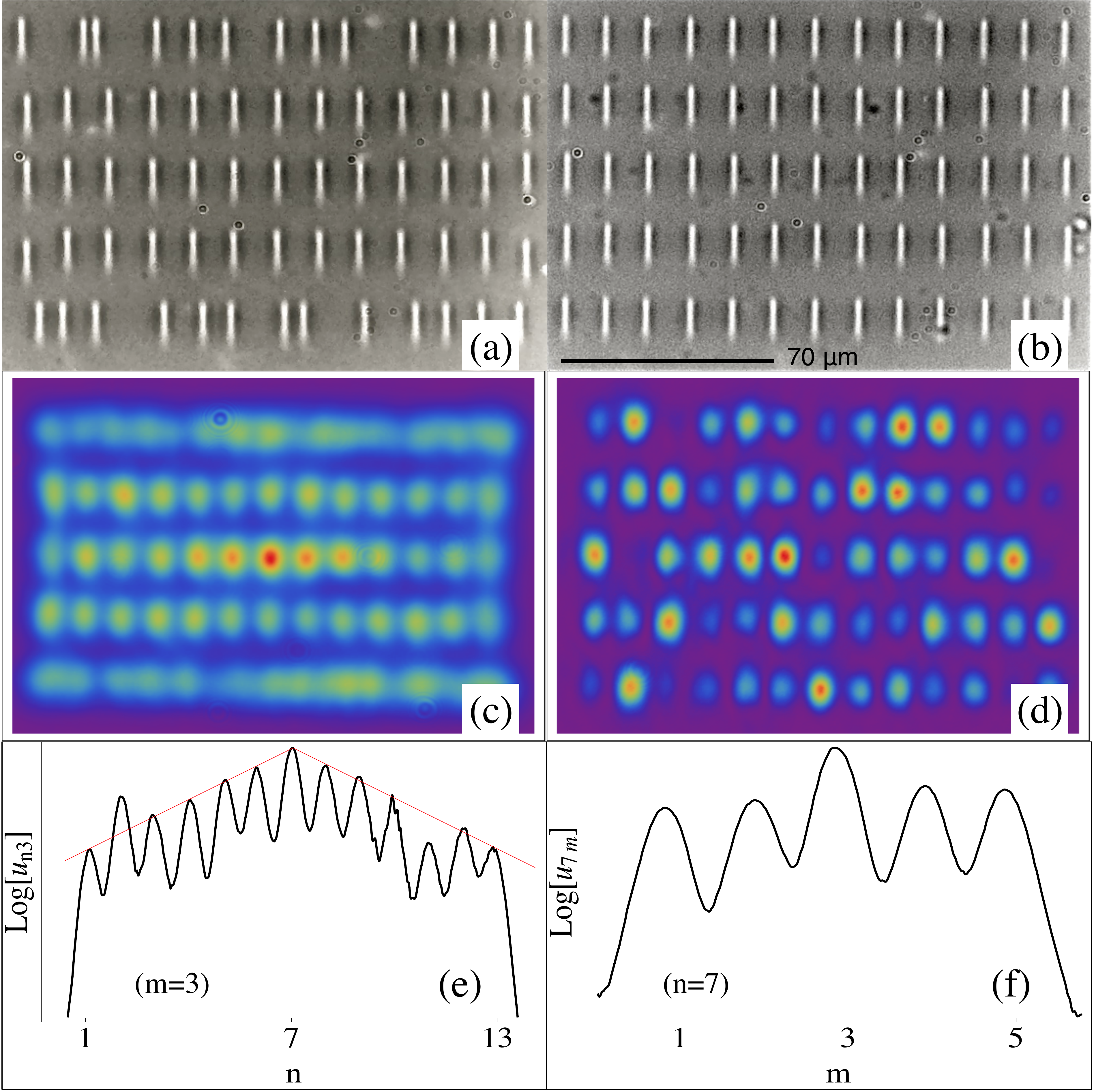}
\caption{Experimental results for the $13\times5$ arrays: (a)
Microscope image of one configuration of the disordered boundary, (b)
ordered array, (c) mean output of  30 realizations of the
disordered boundary waveguide array, $\lambda=840$ $nm$ (d) output of
the ordered array, (e), (f) logarithmic plots of $|u_{n,m_c}|$,
$|u_{n_c,m}|$.} \label{13x5}
\end{figure}

The experimental results, shown in Figs. \ref{5x5} and \ref{13x5},
demonstrate a clear localization tendency around the position of the
input beam, for both, $5\times5$ (Fig. \ref{5x5}) and  $13\times5$
(Fig. \ref{13x5}) arrays. Figure \ref{5x5}(a) shows the output
intensity at the end facet, averaged over 30 different
boundary-disorder realizations. The exponential localization
behavior of the light intensities in the horizontal direction is
shown and clearly seen in Fig. \ref{5x5}(b). Since the global
coupling in the vertical direction is weaker, the decay here is
remarkably stronger (Fig. \ref{5x5}(c)).

A microscopic image of a single realization of the $13\times5$
disordered boundary array is shown in Fig. \ref{13x5}(a), where the
ordered center as well as the boundary layer with disordered spacing
can be seen. For comparison we also show a microscopic image of the
completely ordered $13\times5$ waveguide array. In
Fig. \ref{13x5}(c), the mean output intensity of 30 realizations is
shown, where localization on the central input site is clearly
observed. When comparing with the intensity distribution in an
ordered array (Fig. \ref{13x5}(d)) we find that the propagation
distance is sufficiently long for complete delocalization in the
ordered case. Thus, the observed localization effects can be
attributed to the disordered boundary layer. The stronger coupling
obtained in the $13\times5$ array due to the longer wavelength leads
to better exponential localization in both coupling directions,
shown in Figs. \ref{13x5}(e) and \ref{13x5}(f).

In summary, we have investigated experimentally the influence of a
disordered boundary in a finite 2D coupled waveguide array and found
asymptotic partial localization of the wave packet in the center of
the bulk region far away from the boundary. We conclude that the
presence of a disordered boundary can give rise to Anderson
localization in regions away from the boundary. We conjecture that
this result could be extrapolated to larger finite lattices, if one
allows for a sufficiently long propagation distance.

The authors acknowledge financial support from the German Ministry
of Education and Research (Center for Innovation Competence program,
grant 03Z1HN31), FONDECYT grants 1080374, PFB0824/2008, and a
CONICYT doctoral fellowship.

\pagebreak

\end{document}